\documentclass[a4paper,conference]{IEEEtran}
\IEEEoverridecommandlockouts
\usepackage{hyperref}
\usepackage{cite}
\usepackage{graphicx}

\ifCLASSINFOpdf
\else
\fi
\usepackage{amsmath}
\usepackage{algorithmic}

\usepackage{array}
\usepackage{fixltx2e}

\usepackage{stfloats}
\usepackage{url}

\begin{document}
%
\title{Dual Encoder Fusion U-Net (DEFU-Net) for Cross-manufacturer Chest X-ray Segmentation}

\author{
\IEEEauthorblockN{Lipei Zhang\IEEEauthorrefmark{1}\IEEEauthorrefmark{2},  Aozhi Liu\IEEEauthorrefmark{2}, Jing Xiao\IEEEauthorrefmark{2}, Paul Taylor\IEEEauthorrefmark{1} \thanks{Prof Paul Taylor is corresponding author.}} 
  
\IEEEauthorblockA{\IEEEauthorrefmark{1}University College London, London, UK} 
\IEEEauthorblockA{\IEEEauthorrefmark{2}Ping An Technology (Shenzhen) Co., Ltd., China} 

\IEEEauthorblockA{\IEEEauthorrefmark{1}uceclz0@ucl.ac.uk, p.taylor@ucl.ac.uk}

\IEEEauthorblockA{\IEEEauthorrefmark{2}liuaozhi092@pingan.com.cn, xiaojing661@pingan.com.cn} 
}


%


\maketitle


\begin{abstract}
A number of methods based on deep learning have been applied to medical image segmentation and have achieved state-of-the-art performance. Due to the importance of chest x-ray data in studying COVID-19, there is a demand for state-of-the-art models capable of precisely segmenting soft tissue on the chest x-rays. The dataset for exploring best segmentation model is from Montgomery and Shenzhen hospital which had opened in 2014. The most famous technique is U-Net which has been used to many medical datasets including the Chest X-rays. However, most variant U-Nets mainly focus on extraction of contextual information and skip connections. There is still a large space for improving extraction of spatial features. In this paper, we propose a dual encoder fusion U-Net framework for Chest X-rays based on Inception Convolutional Neural Network with dilation, Densely Connected Recurrent Convolutional Neural Network, which is named DEFU-Net. The densely connected recurrent path extends the network deeper for facilitating contextual feature extraction. In order to increase the width of network and enrich representation of features, the inception blocks with dilation are adopted. The inception blocks can capture globally and locally spatial information from various receptive fields. At the same time, the two paths are fused by summing features, thus preserving the contextual and spatial information for decoding part. This multi-learning-scale model is benefiting in Chest X-ray dataset from two different manufacturers (Montgomery and Shenzhen hospital). The DEFU-Net achieves the better performance than basic U-Net, residual U-Net, BCDU-Net, R2U-Net and attention R2U-Net. This model has proved the feasibility for mixed dataset and approaches state-of-the-art. The source code for this proposed framework is public \url{https://github.com/uceclz0/DEFU-Net}. 
\end{abstract}

%
\IEEEpeerreviewmaketitle

\section{INTRODUCTION}
In the era of pneumonia epidemic in COVID-19, the automatic segmentation of medical images, especially the automatic segmentation of chest X-ray images, has become a key step in the automatic identification and analysis of abnormalities. Accurate and high-performance segmentation models can speed up the clinical workflow and help doctors make more rational decisions for patients. Deep learning is driving the development of medical image segmentation. Compared with the traditional models of computer vision, the deep learning method transcends the limitations of scope of applications \cite{17}. This near-radiologist level achievement of deep learning can also be attributed to the rise to convolutional neural network (CNN). Filter transformation and efficient representation learning are crucial characteristics. Ever since AlexNet \cite{1} has gained huge improvement on the classification on the ImageNet dataset \cite{2}, various convolutional structures have been proposed, such as residual block \cite{3}, densely connected block \cite{24} and inception block \cite{4}. The networks are able to reach deeper and wider range, which is helpful for extraction of low-dimensional and high-dimensional features. In addition, some useful activation functions are helpful for the network to simulate the results generated by human brain, such as ReLU, LeakyReLU, Sigmoid and Softmax. Some efficient optimization algorithms update parameters and accelerate the convergence. For example, stochastic gradient descent (SGD) and Adam optimizer are used in most of the training.

Back to the medical image segmentation, many networks based on CNN make performance approaching the judgement of radiologists. The ground-breaking segmenting network is Fully convolutional network (FCN) \cite{5}. After that, more researchers proposed complex frameworks for improving efficiency of the encoder and decoder. For most medical image datasets, the images are highly similar, unlike the images in ImageNet, which have obvious differences in the edge or shape. For example, the chest X-ray has fuzzy edges and similar areas between normal and abnormal scans. Occasionally, a dataset may be from different X-ray machines or include a small number of images with low quality caused by low contrast, lack of costophrenic angle and biased annotations \cite{18}. Therefore, the limited receptive fields and insufficient context information extracted from FCN may lead to poor performance in some medical datasets. Many more complicated networks have been proposed such as PSPNet \cite{6}, U-Net \cite{7} and DeepLab \cite{8}. They have more sufficient receptive fields and greater ability to obtain richer contextual information, thus obtaining better performance. U-Net is the most classic network in medical image segmentation \cite{19} which is applied widely due to its ability to concatenate contextual information by skipping connections between encoder and decoder. In order to improve the efficiency and accuracy of the network, a number of extensions on U-Net have been proposed. Deep Residual U-Net \cite{9}, which employs residual block into each layer of encoder and decoder, made the network deeper and improved the performance metrics. Recurrent Convolutional layers (RCNN) and Recurrent Residual Convolutional layers (R2CNN) were proposed by Alom et al. \cite{10}, which utilize feature accumulation with the recurrent mechanism. BCDU-Net used bi-directional ConvLSTM instead of skip connection and block of dense convolutions was applied into the bottom encoding layer \cite{11}. Attention mechanism was introduced in skip connections of U-Net\cite{12}. In addition to modifying the network structures, the kernel size has been discussed and explored \cite{20} as well. These models focus on modifying context feature extraction in one path and connection between encoder and decoder. They perform the-state-of-art in some tasks of medical image segmentation. However, they do not focus on the extraction of spatial and contextual information simultaneously, which may cause mis-classification on the pixel of nearby-border. In addition, the different source datasets may lead to the uncertainty of segmentation, and this influence had been discussed in optical coherence tomography \cite{26}. Most networks use a single $3\times3$ convolution kernel convolution kernel, which can not adjust the diversity of object and domain-shifts across manufacturers. Therefore, the model, that can adjust the different device source, deserves to be studied.  

In this paper, we propose a novel extension of U-Net called DEFU-Net to address these problems. A dual-path encoder is constructed to improve the performance of the model. The dual path encoder comprises densely connected recurrent encoding blocks as well as inception encoding blocks with dilation. Some researchers used an inception block to replace the convolution block on each layer in some segmentation task \cite{23}. We employ inception blocks with dilation as the second path to scale up the width of the network. The input features of the second densely connected recurrent block and the first inception block with dilation are shared. The inception block with dilation can adjust both global and local distributions and extract multiple spatial features \cite{4}. It avoids spatial information loss resulted by max-pooling as well. The densely connected recurrent convolution block (DCRC) can facilitate network to extract high-level information and avoid gradient vanishing problem as well\cite{10}. Before concatenation, the extracted information from current DCRC block and inception block with dilation will be fused by summation at each stage. Meanwhile, the fused information will be transferred to the next inception block with dilation to extract spatial features. These processes help the network to obtain more accurate result than other U-Net extensions on the cross-manufacturers dataset from Montgomery and Shenzhen hospital, since our model is able to reduce the influences of the variance among datasets.

\begin{figure*}[ht]
    \centering
    \includegraphics[scale=0.35]{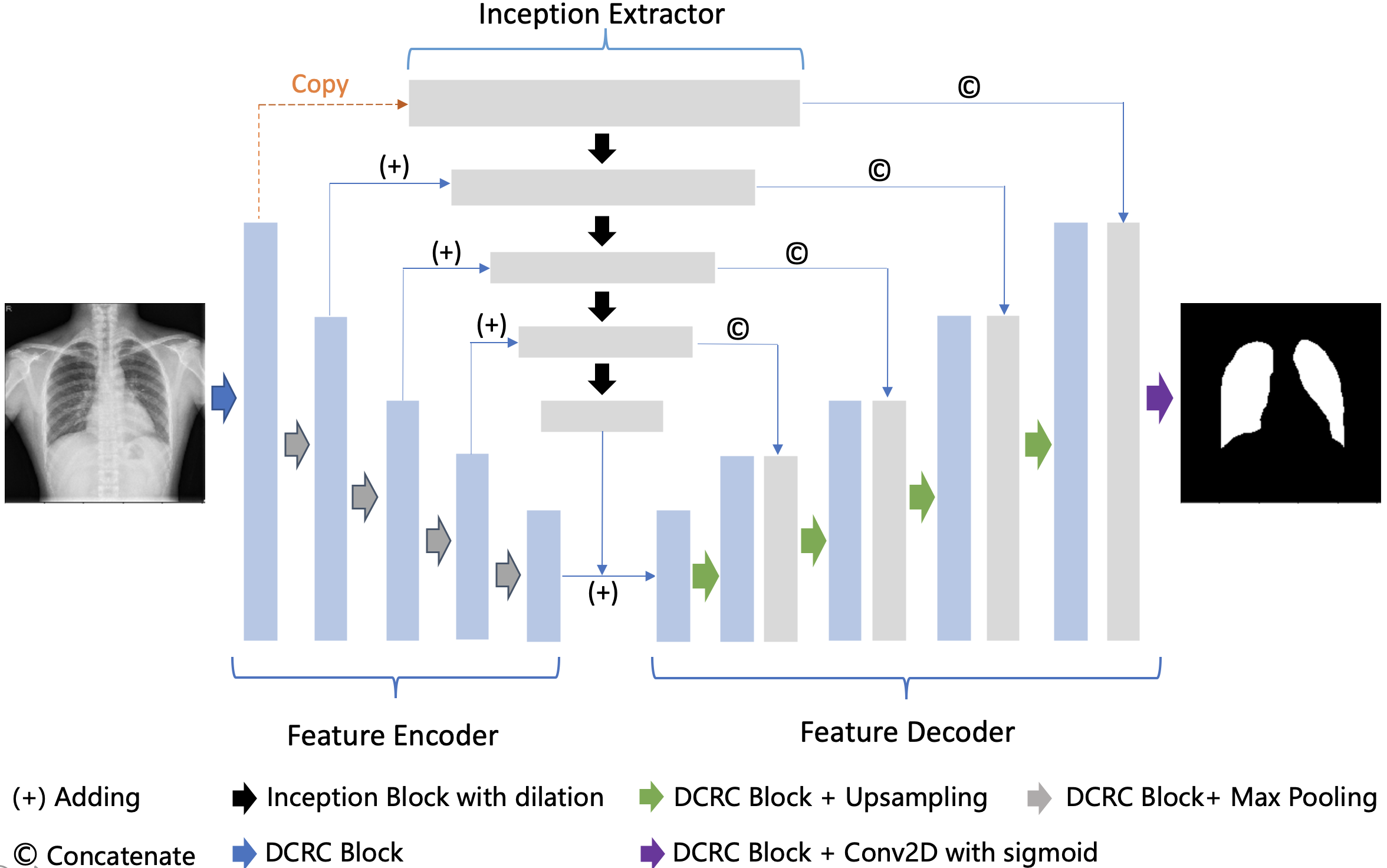}
    \caption{DEFU-Net with Inception dilation Convolution Blocks and Densely Connected Recurrent convolution (DCRC) Blocks} 
\end{figure*}

\begin{figure}[ht]
    \centering
    \includegraphics[scale=0.3]{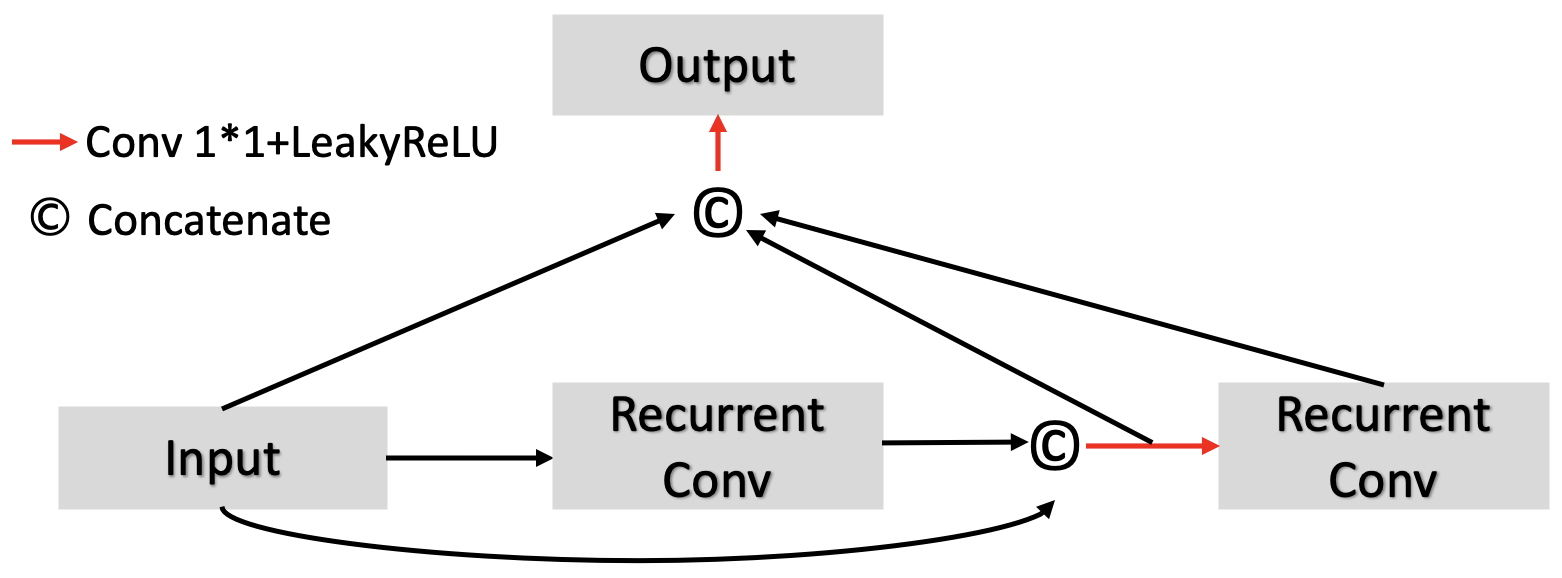}
    \caption{Densely Connected Recurrent convolution (DCRC) Block}
\end{figure}

\begin{figure}[ht]
    \centering
    \includegraphics[scale=0.2]{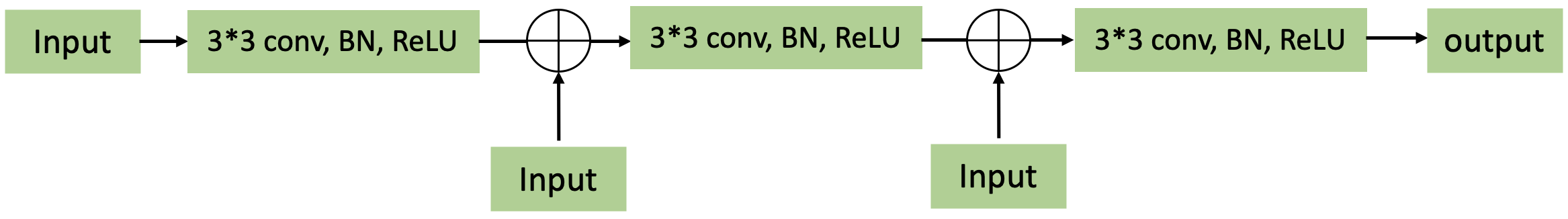}
    \caption{Unfolded Recurrent Convolutional Unit}
\end{figure}

\section{METHOD}
\subsection{Dual Encoder Fusion U-Net}
Inspired by the advantages of U-Net \cite{19}, inception block\cite{4}, DenseNet\cite{24} and recurrent structure\cite{10}, the dual encoder fusion U-Net is proposed, as shown in Fig. 1. This new framework follows the classic encoding and decoding structure of U-Net. The blue boxes represent the group of feature maps in each layer and the grey boxed are the set of feature maps from inception extracting path. On each stage of the encoder, we apply two recurrent blocks with densely connections and max-pooling (grey arrow). After feature maps extracted by first DCRC blocks (blue arrow), they will be copied to inception extractor with dilation (black arrow). In first stage of inception extractor, $X_1 = Y_1$ and output $Y_2$ will be equal to $E_{1}(Y_1)$, where $E_{n}$ indicates the process of feature extraction from inception block with dilation. $X_n$ and $Y_n$ denote input feature maps from feature encoder and inception extractor in $n^{th}$ stage, respectively. The information extracted by DCRC block and inception block will be fused by pixel-wisely summation (adding (+)) in the rest of the encoder. Each set of fused features will serve as an input to the next inception block with dilation. $D_n$ represents the process of feature extraction from DCRC. This process can be formulated:

\begin{equation}
    \begin{split}
        & X_{n+1} = D_{n}(X_n) \\
        & Y_{n+1} = E_{n}(X_n + Y_n)
    \end{split}
\end{equation}

These operations can enrich spatial and context features. The fused feature maps $(X_n + Y_n)$ are prepared for concatenating to decoder stages accordingly. The concatenations encourages the information reused in decoder. In order to avoid increasing parameters, Up-sampling is adopted in decoding part. The green arrow indicates DCRC block + Up-sampling. The up-sampling is beneficial to recover boundaries location from low-dimensional features. Moreover, we modified the number of filters in the bottom layer, as same as $4^{th}$ layer. This modification can reduce computational budget and avoid yielding useless feature maps.

\subsection{Densely Connected Recurrent Block}

The densely connected recurrent convolution block (DCRC block) in our network is inspired by R2 block proposed by Altom et al.\cite{10} and DenseNet\cite{24}. The unfolded structure is shown in Fig. 2. The recurrent unit can help feature accumulation and extract useful information precisely. The multiple kernels will extract information from accumulated feature maps. The structure of recurrent convolutional unit is illustrated in Fig. 3. In order to improve the stability of training, batch normalization is used in the block \cite{21}. Meanwhile, the block includes the densely connected mechanism. After each recurrent unit, the number of channels will be increased with densely connection mechanism because the output and all of previous features are concatenated. With limited computational resource, we choose Conv($1\times1$)-LeakyReLU for recover the number of channels, which is similar to the bottleneck layers used in DenseNet \cite{24} to reduce the number of channels. The multi-connections can enforce individual layer obtain deep supervision additionally from loss function \cite{25}. The network can become deeper and the convergence is faster in the training process.

\subsection{Inception extractor with dilation}
\begin{figure}
    \centering
    \includegraphics[scale=0.4]{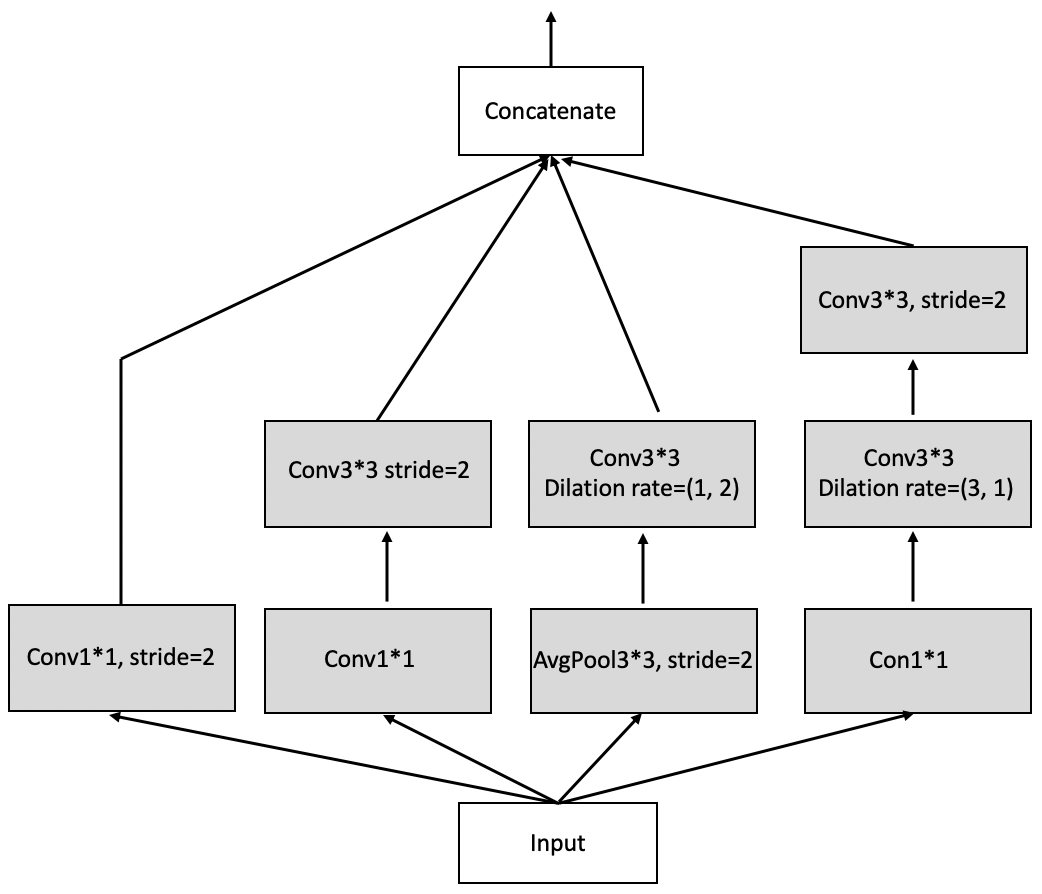}
    \caption{Inception block with dilation}
\end{figure}
In our network, we take the classical Inception V2 and V3 \cite{16} as a reference. In X-ray scanning, the height of lungs is usually greater than the width. Inspired by the success of atrous/dilated convolution\cite{27} in medical image segmentation\cite{28}, Conv($3\times3)$ with dilation rate $(3, 1)$ and $(1, 2)$ are introduced in this block for expand receptive fields on height and width respectively. The details of structure are shown in Fig. 4. This structure uses three ways to decrease dimension, including $1\times1$ convolution, $3\times3$ convolution with stride 2 and $3\times3$ average-pooling. In order to contain convolutional continuity and realize down-sampling, the $3\times3$ kernel with stride 2 can help preserve spatial characteristics and avoid information loss directly caused by max-pooling. The $1\times1$ kernel with stride 2 can enhance the non-linear capacity \cite{16}. Especially, two branches include dilated convolutions. Based on the equations $H_{dilated}/W_{dilated} = (Dilation Rate-1)\times(Kernel Size-1)+Kernel Size$, the new kernel size will be $7\times3$ and $3\times5$ respectively. Thus, dilated receptive fields on height is larger than dilated receptive fields on width for adjusting difference of learning on shape. The correlation and continuity of high-level semantic features in height or width will be learned. By combining with other kernels and average-pooling, this modified inception can aggregate multi-scale contextual information for dense prediction architectures, thus improving the performance\cite{29}. From the second layer to the bottom of encoder, the feature extracted by the last inception block and the feature extracted by DCRC block will be fused by summation and they will be concatenated to the decoder. The summation on the element-wise feature has been proved to have a great performance outside U-Net \cite{14}. The rich spatial and context features are integrated into the decoding part.  

\section{EXPERIMENTS}
\subsection{Dataset}
We mainly evaluated the DEFU-Net on chest X-ray from Montgomery and Shenzhen hospital which opened in 2014 \cite{13}. This dataset contains many diagnoses such as infiltration, fibrosis, pneumonia and tuberculosis. These diseases have similar radiogram and we mainly focus on segmenting lung soft tissue. Therefore, we use this dataset to investigate a state-of-the-art pre-trained model. This dataset includes chest X-ray scans from two manufacturers. 138 patient's images from Montgomery and 566 patient's images from Shenzhen are applied in this segmentation task respectively. The total number of normal lung was 359, while the number of abnormal lung was 345. The size of X-ray from Montgomery Country is either with $4020\times4892$ or $4892\times4020$ pixels. The size of Shenzhen chest X-ray is $3K\times3K$. The pixel-wise lung mask annotations are offered in the two datasets. Specially, X-ray scans from Montgomery are annotated in left and right lung respectively. Thus, we combined left and right lung segmentation masks from Montgomery and resized all the X-ray scans from two dataset to $512\times512$ pixels. All scans were transformed to a single channel as grey-scale. All masks were dilated to gain more information on the edge of lungs for training. After pre-processing, the data set was mixed and divided into 528 training set, 76 images for validation set and 100 for testing set randomly. To further explore the robustness of the proposed model, we trained models from one hospital and tested it in another hospital. (1. Training on Montgomery (M) and testing on shenzhen (S); 2. Training on Shenzhen (S) and testing on Montgomery (M)). In the training process, the training data were augmented by rotation, shifting, shearing, zooming and flipping in order to avoid over-fitting \cite{1}. 

\begin{figure*}[ht]
    \centering
    \includegraphics[scale=0.29]{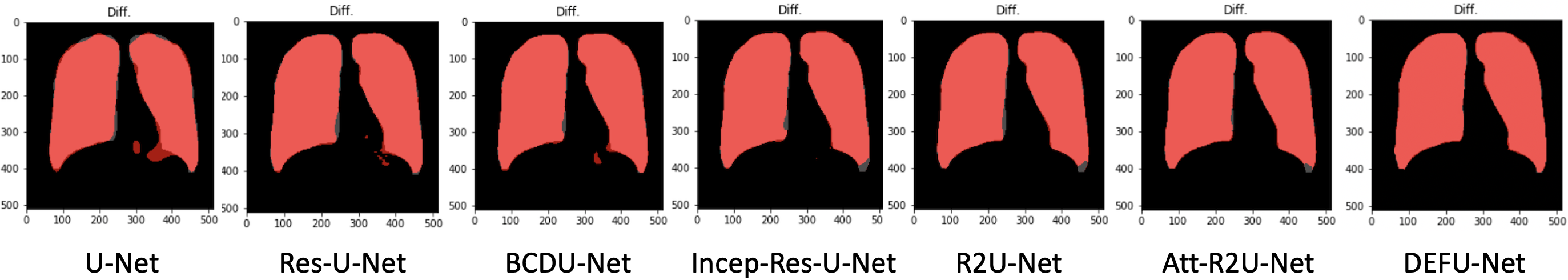}
    \caption{Comparison between Prediction and Ground Truth of Shenzhen X-Ray in Variant U-Nets. Grey color: ground truth. Red color: prediction. The model trained in mixed dataset is used here.} 
\end{figure*}

\begin{figure*}[ht]
    \centering
    \includegraphics[scale=0.3]{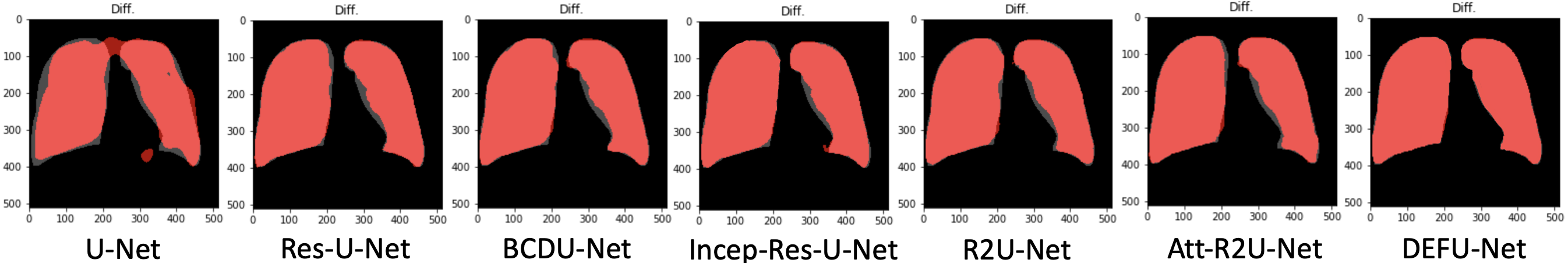}
    \caption{Comparison between Prediction and Ground Truth of Montegomery X-Ray in Variant U-Nets. Grey color: ground truth. Red color: prediction. The model trained in mixed dataset is used here.} 
\end{figure*}

\subsection{Training of the neural network}
The training environment was based on Keras 2.2.4 with Tensorflow 1.13 backend. GPU was 1080Ti. The batch size was set to 2. The input and output were $512\times512\times1$. We used classic dice loss Eq.2 for semantic segmentation and Adam optimizer \cite{22} with default parameters. 
\begin{equation}
    L_{dice}= -\frac{2\sum\limits_{i=1}^{N}g_{i}p_{i}+1}{\sum\limits_{i=1}^{N}g_i^{2} + \sum\limits_{i=1}^{N}p_i^{2}+1}
\end{equation}
Where $g_{i}\in{0, 1}$ represents the ground truth on the pixel level. $p_{i}\in[0, 1]$ refers to the probability value of the prediction on each pixel. $N$ is the total number of the pixel. Adding 1 is for preventing divided by zero. 

The “reduce learning rate on plateau" with initial learning rate $1\times10^{-5}$, factor = 0.2 and patience = 5) and ”Early Stopping” (patience = 5) were applied in training. The total training epochs was set to 175. Meanwhile, In order to understand the capacity of inception extractor with dilation, we combine it with Res-U-Net. The U-Net \cite{7}, Residual U-Net (Res-U-Net) \cite{9}, Incept-Res-U-Net, BCDU-Net \cite{11}, R2U-Net and R2-Att-U-Net, DEFU-Net had all been trained on our dataset.

\subsection{Evaluation approaches}
We used 7 evaluation metrics for our training and testing predictions, including binary accuracy (AC), dice coefficient (Dice Coef = -dice loss), intersection over union (IOU), precision, recall, F1-score and area under curve (AUC). The Dice Coef and IOU can be expressed in the following equations, where GT represent the ground truth and PR refers to the prediction result.
\begin{equation}
    dice = \frac{2 \times|GT|\bigcap PR|+1}{|GT|+|PR|+1}
\end{equation}
\begin{equation}
    JS (IOU) = \frac{|GT \bigcap PR|+1}{|GT \bigcup PR|+1}
\end{equation}
In order to calculate AC, precision, recall, F1-score, we need to employ True Positive (TP), True Negative (TN), False Positive (FP), and False Negative (FN). The metrics can be obtain by following equations. The AUC can be calculated by receiver operating characteristic curve. 
\begin{equation}
    AC = \frac{TP+TN}{TP+TN+FP+FN}
\end{equation}
\begin{equation}
    Precision = \frac{TP}{TP+FP}
\end{equation}
\begin{equation}
    Recall = \frac{TP}{TP+FN}
\end{equation}
\begin{equation}
    F1 = 2 \times \frac{Precision \times Recall}{Precision + Recall}
\end{equation}

\begin{table*}[htb]
	\renewcommand{\arraystretch}{1.3}
	\centering
	\caption{Training Result}	
	\begin{tabular}{llllllll}
		\hline
		
		\textbf{Training}& Dice &  AC & IOU& Precision& Recall & 	F1 Score & 	AUC
		\\	\hline	
		U-Net & 
		0.9039
		& 	
		0.9403
		&	
		0.9695
		&
		0.9063
		&	
		0.9032
		& 
		0.9043
		& 
		0.8757
		\\

		Res-U-Net
		& 
		0.9688
		& 
		0.9807
		& 
		0.9902
		&
		0.9744
		& 
		0.9641
		& 
		0.9689
		& 
		0.9748
		\\

		BCDU-Net
		& 
		0.9707
		& 
		0.9816
		& 
		0.9908
		&
		0.9761
		& 
		0.9665
		& 
		0.9707
		& 
		0.9719
		\\
		
		Incep-Res-U-Net
		&
		0.9740
		&
		0.9837
		&
		0.9917
		&
		0.9785
		&
		0.9704
		&
		0.9741
		&
		0.9785
		\\

		R2-UNet
		& 
		0.9793
		& 
		0.9868
		&
		0.9932
		&
		0.9823
		& 
		0.9771
		& 
		0.9795
		& 
		0.9800
		\\

		Att-R2U-Net
		& 
		0.9791
		& 
		0.9867
		& 
		0.9931
		&
		0.9825
		& 
		0.9766
		& 
		0.9793
		& 
		0.9807
		\\
	    DEFU-Net
		&
		\textbf{0.9858}
		&
		\textbf{0.9910}
		&
		\textbf{0.9954}
		&
		\textbf{0.9871}
		&
		\textbf{0.9848}
		&
		\textbf{0.9859}
		&
		\textbf{0.9828}
		\\

		\hline
	\end{tabular}
\end{table*}
\vspace{-1cm}

\begin{table*}[htb]
	\renewcommand{\arraystretch}{1.3}
	\centering
	\caption{Validation Result}
	
	\begin{tabular}{llllllll}
		\hline
		
		\textbf{Validation} 
		&
		Dice
		& 
		AC
		& 
		IOU
		&
		Precision
		& 
		Recall
		& 
		F1 Score
		& 
		AUC
		\\
		\hline	
		U-Net
		& 
		0.7751
		& 
		0.8801
		& 
		0.9396
		&
		0.8210
		& 
		0.7350
		& 
		0.7753
		& 
		0.8763
		\\
		
		Res-U-Net
		& 
		0.9167
		& 
		0.9541
		& 
		0.9769
		&
		0.9336
		& 
		0.9016
		& 
		0.9168
		& 
		0.9748
		\\
		
		BCDU-Net
		& 
		0.9482
		& 
		0.9720
		&
		0.9860
		&
		0.9617
		& 
		0.9370
		& 
		0.9482
		& 
		0.9720
		\\
		
		Incep-Res-U-Net
		&
		0.9495
		&
		0.9722
		&
		0.9860
		&
		0.9730
		&
		0.9298
		&
		0.9496
		&
		0.9788
		\\
		
		R2U-Net
		& 
		0.9530
		& 
		0.9750
		& 
		0.9874
		&
		0.9721
		& 
		0.9365
		& 
		0.9532
		& 
		0.9800
		\\	
		Att-R2U-Net
		& 
		0.9497
		& 
		0.9747
		& 
		0.9872
		&
		0.9706
		& 
		0.9339
		& 
		0.9498
		& 
		0.9809
		\\
		DEFU-Net
		&
		\textbf{0.9578}
		&
		\textbf{0.9765}
		&
		\textbf{0.9882}
		&
		\textbf{0.9739}
		&
		\textbf{0.9440}
		&
		\textbf{0.9579}
		&
		\textbf{0.9826}
		\\

		\hline
	\end{tabular}	
\end{table*}
\vspace{-1cm}

\begin{table*}[htb]
	\renewcommand{\arraystretch}{1.3}
	\centering
	\caption{Testing Result}
	\begin{tabular}{llllllll}
		\hline
		
		\textbf{Testing}
		& 
		Dice
		& 
		AC
		& 
		IOU
		&
		Precision
		& 
		Recall
		& 
		F1 Score
		& 
		AUC
		\\
		\hline
		
		U-Net
		& 
		0.9190
		& 
		0.9539
		&
		0.9764
		& 
		0.9445
		& 
		0.9013
		& 
		0.9194
		& 
		0.9556
		\\
		Res-U-Net
		& 
		0.9602
		& 
		0.9767
		& 
		0.9880
		&
		0.9690
		& 
		0.9541
		& 
		0.9605
		& 
		0.9804
		\\
		BCDU-Net
		& 
		0.9658
		& 
		0.9795
		& 
		0.9897
		&
		0.9645
		& 
		0.9684
		& 
		0.9658
		& 
		0.9813
		\\
		Incep-Res-U-Net
		&
		0.9616
		&
		0.9775
		&
		0.9885
		&
		0.9714
		&
		0.9545
		&
		0.9620
		&
		0.9802
		\\
		R2U-Net
		& 
		0.9647
		& 
		0.9794
		& 
		0.9896
		&
		0.9748
		& 
		0.9568
		& 
		0.9648
		& 
		0.9815
		\\
		Att-R2U-Net
		& 
		0.9644
		& 
		0.9793
		& 
		0.9895
		&
		0.9732
		& 
		0.9579
		& 
		0.9645
		& 
		0.9823
		\\
		DEFU-Net
		&
		\textbf{0.9667}
		&
		\textbf{0.9804}
		&
		\textbf{0.9901}
		&
	    0.9731
		&
		0.9619
		&
		\textbf{0.9667}
		&
		0.9816
		\\
	
		\hline
	\end{tabular}
\end{table*}
\vspace{-0cm}

\begin{table*}[htb]
	\renewcommand{\arraystretch}{1.3}
	\centering
	\caption{Test Result without Mixed Dataset. 1. Training on Montgomery (M) and testing on shenzhen (S); 2. Training on Shenzhen (S) and testing on Montgomery (M)}
	\begin{tabular}{llllllll}
		\hline
		\textbf{Test Dice/F1 Score}
		& 
		U-Net
		& 
		Res-U-Net
		& 
		BCDU-Net
		& 
		Incep-Res-U-Net
		&
		R2U-Net
		& 
		Att-R2U-Net
		&
		DEFU-Net
		\\
		\hline
		Train: M/Test: S
		&
		0.8556/0.8568
		&
		0.6642/0.6653 
		&
	    0.9088/0.9089
		&
		0.8896/0.8918 
		&
		0.9122/0.9141
		&
		0.9069/0.9097 
		&
		\textbf{0.9154/0.9158}
		\\
		Train: S/Test: M
		&
		0.7706/0.7739 
		&
		0.8162/0.8191 
		&
		0.7671/0.9082
		&
		0.8984/0.9022 
		&
		0.8659/0.8713 
		&
		0.8659/0.8713
		&
		\textbf{0.9227/0.9231}
		\\
		\hline
	\end{tabular}
\end{table*}
\vspace{2cm}



\section{RESULTS}


For comparison, the evaluation metrics of training, validation and testing are shown in Tables 1, 2 and 3. DEFU-Net generates the highest Dice, AC, IOU, precision, recall, F1-score and AUC in training, validation dataset. In testing dataset, the DEFU-Net still outperforms most of the other metrics. Meanwhile, under the "Early Stopping' mechanism, U-Net, Res-U-Net, BCDU-Net and Incep-Res-U-Net stop training at $70^{th}$ epoch, $50^{th}$ epoch, $72^{th}$ and $60^th$ respectively. The R2U-Net and Att-R2U-Net stop at $100^{th}$. Our model can train 145 epochs with higher metrics. The AC of our model can reach $0.9776$ after two epochs. They demonstrate that our model has fast convergence and fits our cross-manufacturer data better in training. Moreover, we can see that the inception path with dilation can boost performance beyond U-Net, Res-U-Net and BCDU-Net. It is shown that this path are actually effective for our cross-manufacturer segmentation task. After combining the inception blocks with dilation and densely connection recurrent blocks as dual path encoder, the dataset can be fit with the best performance. 

In the experiments of training and testing without mixed dataset, the performance of models is mainly compared by testing dice and F1 Score. In Table 4, the proposed DEFU-Net have highest metrics in the both processes of Train: M/Test: S and Train: S/Test: M. The DEFU-Net is robust and convincing for segmentation task of data across manufacturers. 

We visualise the difference between prediction and ground truth of Montgomery and Shenzhen dataset in Fig. 5 and Fig. 6. The grey region indicates the ground truth and red color represents predicted pixel. The model trained in mixed dataset is used here. From left to right, the predictions fit the ground truth gradually. In order to solve to the adaptability problem on diversity of lung size, shape and position, the model needs to learn these differences. From the comparison diagrams, they show the advantages of inception with dilation and DCRC block clearly. This extractor path improves classification on the pixel of nearby-border and the DCRC blocks helps reduce the rate of True Negative and the False Positive. The DEFU-Net achieves superior performance. 

\section{DISCUSSION AND CONCLUSION}
Against the backdrop of COVID-19, the state-of-the-art pre-trained model about segmentation is significant for the future COVID-19 chest X-ray to diagnose pneumonia. In this paper, we proposed an innovative network structure called DEFU-Net to segment the opened cross-manufacturer chest X-ray dataset with great performance. We applied a dual path framework to enrich the features extracted from the encoder. The inception path with dilation can help the model to capture spatial information with multi-scale kernels and increase the width of the network. The densely connected recurrent block increases the depth of the network. Information from low-level to high-level can be captured. The pixel-wise summations of features from two paths preserve more optimal information during decoding. Meanwhile, we change the number of feature maps in bottom layer, which is same with $4^{th}$ level. The demand for computational space will be halved. On this combination of two public datasets, the DEFU-Net have better fitting ability on the segmentation of edges and small areas. We observe our model performing state-of-the-art compared to the aforementioned model. In the future, our model may carry out experiments of segmentation tasks under cross-manufacturer COVID-19 pneumonia X-ray, and we may explore the feasibility of expanding to 3D images for more complex segmentation.

\bibliographystyle{abbrv}
\bibliography{IEEEabrv, root.bib}

\begin{thebibliography}{10}

\bibitem{10}
M.~Z. Alom, M.~Hasan, C.~Yakopcic, T.~M. Taha, and V.~K. Asari.
\newblock Recurrent residual convolutional neural network based on u-net
  (r2u-net) for medical image segmentation.
\newblock {\em CoRR}, abs/1802.06955, 2018.

\bibitem{11}
R.~Azad, M.~Asadi-Aghbolaghi, M.~Fathy, and S.~Escalera.
\newblock Bi-directional convlstm u-net with densley connected convolutions,
  2019.

\bibitem{23}
D.~E. Cahall, G.~Rasool, N.~C. Bouaynaya, and H.~M. Fathallah-Shaykh.
\newblock Inception modules enhance brain tumor segmentation.
\newblock {\em Frontiers in Computational Neuroscience}, 13:44, 2019.

\bibitem{8}
L.~Chen, G.~Papandreou, I.~Kokkinos, K.~Murphy, and A.~L. Yuille.
\newblock Deeplab: Semantic image segmentation with deep convolutional nets,
  atrous convolution, and fully connected crfs.
\newblock {\em CoRR}, abs/1606.00915, 2016.

\bibitem{27}
L.-C. Chen, Y.~Zhu, G.~Papandreou, F.~Schroff, and H.~Adam.
\newblock Encoder-decoder with atrous separable convolution for semantic image
  segmentation, 2018.

\bibitem{2}
J.~Deng, W.~Dong, R.~Socher, L.~Li, K.~Li, and F.~Li.
\newblock Imagenet: {A} large-scale hierarchical image database.
\newblock In {\em 2009 {IEEE} Computer Society Conference on Computer Vision
  and Pattern Recognition {(CVPR} 2009), 20-25 June 2009, Miami, Florida,
  {USA}}, pages 248--255. {IEEE} Computer Society, 2009.

\bibitem{26}
J.~Fauw, J.~Ledsam, B.~Romera-Paredes, S.~Nikolov, N.~Tomasev, S.~Blackwell,
  H.~Askham, X.~Glorot, B.~O’Donoghue, D.~Visentin, G.~Driessche,
  B.~Lakshminarayanan, C.~Meyer, F.~Mackinder, S.~Bouton, K.~Ayoub, R.~Chopra,
  D.~King, A.~Karthikesalingam, and O.~Ronneberger.
\newblock Clinically applicable deep learning for diagnosis and referral in
  retinal disease.
\newblock {\em Nature Medicine}, 24, 09 2018.

\bibitem{3}
K.~He, X.~Zhang, S.~Ren, and J.~Sun.
\newblock Deep residual learning for image recognition.
\newblock {\em CoRR}, abs/1512.03385, 2015.

\bibitem{24}
G.~Huang, Z.~Liu, L.~van~der Maaten, and K.~Q. Weinberger.
\newblock Densely connected convolutional networks, 2016.

\bibitem{21}
S.~Ioffe and C.~Szegedy.
\newblock Batch normalization: Accelerating deep network training by reducing
  internal covariate shift, 2015.

\bibitem{13}
S.~Jaeger, S.~Candemir, S.~Antani, Y.-X.~J. Wáng, P.-X. Lu, and G.~Thoma.
\newblock Two public chest x-ray datasets for computer-aided screening of
  pulmonary diseases.
\newblock {\em Quantitative Imaging in Medicine and Surgery}, 4(6), 2014.

\bibitem{14}
B.~Kayalibay, G.~Jensen, and P.~van~der Smagt.
\newblock Cnn-based segmentation of medical imaging data, 2017.

\bibitem{22}
D.~P. Kingma and J.~Ba.
\newblock Adam: A method for stochastic optimization, 2014.

\bibitem{1}
A.~Krizhevsky, I.~Sutskever, and G.~E. Hinton.
\newblock Imagenet classification with deep convolutional neural networks.
\newblock In {\em Proceedings of the 25th International Conference on Neural
  Information Processing Systems - Volume 1}, NIPS’12, page 1097–1105, Red
  Hook, NY, USA, 2012. Curran Associates Inc.

\bibitem{17}
Y.~LeCun, Y.~Bengio, and G.~Hinton.
\newblock Deep learning.
\newblock {\em Nature}, 521:436--44, 05 2015.

\bibitem{25}
C.-Y. Lee, S.~Xie, P.~Gallagher, Z.~Zhang, and Z.~Tu.
\newblock Deeply-supervised nets, 2014.

\bibitem{5}
J.~Long, E.~Shelhamer, and T.~Darrell.
\newblock Fully convolutional networks for semantic segmentation.
\newblock {\em CoRR}, abs/1411.4038, 2014.

\bibitem{12}
O.~Oktay, J.~Schlemper, L.~L. Folgoc, M.~Lee, M.~Heinrich, K.~Misawa, K.~Mori,
  S.~McDonagh, N.~Y. Hammerla, B.~Kainz, B.~Glocker, and D.~Rueckert.
\newblock Attention u-net: Learning where to look for the pancreas, 2018.

\bibitem{20}
C.~Peng, X.~Zhang, G.~Yu, G.~Luo, and J.~Sun.
\newblock Large kernel matters -- improve semantic segmentation by global
  convolutional network, 2017.

\bibitem{7}
O.~Ronneberger, P.~Fischer, and T.~Brox.
\newblock U-net: Convolutional networks for biomedical image segmentation.
\newblock {\em CoRR}, abs/1505.04597, 2015.

\bibitem{19}
B.~A. Skourt, A.~E. Hassani, and A.~Majda.
\newblock Lung ct image segmentation using deep neural networks.
\newblock {\em Procedia Computer Science}, 127:109--113, 2018.

\bibitem{4}
C.~Szegedy, W.~Liu, Y.~Jia, P.~Sermanet, S.~E. Reed, D.~Anguelov, D.~Erhan,
  V.~Vanhoucke, and A.~Rabinovich.
\newblock Going deeper with convolutions.
\newblock {\em CoRR}, abs/1409.4842, 2014.

\bibitem{16}
C.~Szegedy, V.~Vanhoucke, S.~Ioffe, J.~Shlens, and Z.~Wojna.
\newblock Rethinking the inception architecture for computer vision, 2015.

\bibitem{18}
Y.~Tang, Y.~Tang, J.~Xiao, and R.~M. Summers.
\newblock Xlsor: A robust and accurate lung segmentor on chest x-rays using
  criss-cross attention and customized radiorealistic abnormalities generation,
  2019.

\bibitem{29}
F.~Yu and V.~Koltun.
\newblock Multi-scale context aggregation by dilated convolutions, 2015.

\bibitem{9}
Z.~Zhang, Q.~Liu, and Y.~Wang.
\newblock Road extraction by deep residual u-net.
\newblock {\em CoRR}, abs/1711.10684, 2017.

\bibitem{6}
H.~Zhao, J.~Shi, X.~Qi, X.~Wang, and J.~Jia.
\newblock Pyramid scene parsing network.
\newblock {\em CoRR}, abs/1612.01105, 2016.

\bibitem{28}
X.-Y. Zhou, J.-Q. Zheng, P.~Li, and G.-Z. Yang.
\newblock Acnn: a full resolution dcnn for medical image segmentation, 2019.

\end{thebibliography}

\end{document}